# Macroscopic effects of tunneling barriers in nanotube bundles aggregates

M. Salvato[a], M. Cirillo[a], M. Lucci[a], S. Orlanducci[b], I. Ottaviani[a], M.L. Terranova[b], and F. Toschi[b]

[a)] *Dipartimento di Fisica and MINAS Laboratory, Università di Roma "Tor Vergata",I- 00133Roma*

[b)] *Dipartimento di Scienze e Tecnologie Chimiche and MINAS Laboratory, Università di Roma "Tor Vergata" , I- 00133 Roma*

## Abstract

We report on experiments conducted on single walled carbon nanotube bundles aligned in chains and connected through a natural contact barrier. The dependence upon the temperature of the transport properties is investigated for samples having different characteristics. Starting from two bundles separated by one barrier deposited over four contact probes, we extend the study of the transport properties to samples formed by chains of several bundles. The systematic analysis of the properties of these aggregates shows the existence of two conduction regimes in the barrier. We show that an electrical circuit taking into account serial and parallel combinations of voltages generated at the junctions between bundles can model the samples consistently.

The interest for carbon nanotubes (CNT) in electronic industry has much grown after the fabrication of transistors and diodes based on metal or semiconducting CNT [1]. Although devices based on single CNT have been proposed and analyzed in several contexts [2] it is predictable that aggregates of CNTs could also offer interesting counterparts for applications. The properties of the aggregates might be more complex to handle and to interpret with respect to the isolated nanotubes, however, it is the purpose of this paper to show that Single-Walled Carbon Nanotube (SWCNT) bundles systems have properties which can be well understood in terms of physical and transport phenomena. We report indeed on transport measurements performed on bundles containing semiconducting and metallic SWCNT aligned along the direction of an external bias current. We characterize systematically the temperature dependence of the transport properties of the samples and analyze the observed effects in terms of the topological properties. Measurements of the current-voltage characteristics are interpreted in terms of the junctions between the bundles in the aggregates and the effects due to the series connection along the chains of bundles are clearly demonstrated.

Bundles of semiconducting and metallic SWCNT were deposited on insulating $SiO_2$ substrates where metallic Au contacts had been previously patterned with four lead configurations as shown in the inset of Fig.1a. We call this contact configuration NTPR1 (Nano Tube PRobe 1) in order to distinguish it from another configuration that shall be later described; in any case, all the transport measurements herein presented are performed by a four probe technique. The voltage probes distance is fixed at 5μm and the bundles are aligned along the current direction by a dielectrophoretic technique described elsewhere [3]. The length of each bundle ranges between 2μm and 3μm and their diameter is typically of the order of 100nm; thus, for this probing configuration even a chain formed by few bundles (in principle even two) can connect the voltage electrodes. The Scanning Electron Microscopy (SEM) shown in Fig.1a, performed on one of the samples, confirms the expected length and diameter of the bundles with the formation of a contact junction (indicated by the arrow) between two of them. This junction is supposed to be formed by the action of van der Waals forces between the graphenic surfaces with a consequent potential barrier formation [4].

Transport measurements have been performed in a high vacuum cryocooler whose nominal cooling power at the cold finger where the samples are attached is 0.5W at $T$=4.2K. A resistance vs. temperature ($R$-$T$) dependence of the sample of Fig.1a in the NTPR1 contact configuration, measured with a bias current of 10nA, is reported in the upper inset of Fig.1b. The sharp increase at temperature lower than 30K is due to the semiconducting nature of the sample. The line is a fit to the data obtained by the Fluctuation Induced Tunnel (FIT) [5] model based on a tunnel mechanism of the charge carriers across a potential barrier enhanced by thermal fluctuations. In the case of CNT, this barrier is assumed to be formed at the connection between the bundles [6], shown by the arrow in Fig.1a for our sample. The FIT model predicts a temperature dependence of the electrical resistance given by $R = R_0 e^{T_1/(T+T_2)}$ where $T_1$ is proportional to the potential barrier, $T_2$ is the temperature below which FIT regime is active and $R_0$ is the resistance at a given temperature. In our case $T_1$=204K and $T_2$=30K give the best fit of the $R$-$T$ data in agreement with previous results[6].

The current-voltage ($I$-$V$) characteristics measured at $T$=5K are reported in the main panel of Fig.1b. The curve is rather symmetrical showing that neither metallic contacts nor (insulating) substrate influence the Fermi energy level position with respect to the conduction and valence bands inside the bundles [7]; the nonlinearity in this current-voltage characteristic was not present at room temperature and we could record it only below 80*K*. By the measured resistance at low bias ($I$=10*nA*) and the bundles dimensions measured by SEM we deduced a resistivity of our sample formed by two connected bundles of 2.3kΩ·μm at $T$=5*K* and, using the data of upper inset of Fig.1b, 21Ω·μm at $T$=240K.

Plotting the data in double logarithmic scale, as reported in the lower inset of Fig.1b, more insight can be gained. The I-V characteristic shows that two straight lines with different slope can fit the data in the low and high bias current region. The crossing of the two straight lines at voltage of $V^*$=140mV and current $I^*$=0.1μA provide an indication on the separation between the two different transport regimes. Both the slopes are fitted with a power function $I=V^{\alpha}$ with the exponent $\alpha$=1.1 and 1.7 respectively for the low and high bias current part of the characteristic respectively. We observed the same properties on dozens of samples with the only difference that increasing the distance between the voltage electrodes, the *I-V* characteristic of Fig.1b scaled up in voltage. For this reason we decided to change the probe design in order to investigate systematically the properties of the aggregates upon the distance between the voltage probing electrodes.

We stepped then to the contacts pattern NTPR2 shown in the inset of Fig. 2a: here we placed "inner" electrodes, 100μm length, at multiples of 20μm while the leftmost and rightmost contact pads were used for biasing the aggregates which are deposited all over the contact pattern and aligning them along the direction orthogonal to the electrodes. We could probe the voltages of the aggregates at distances of 20,40,60, and 80μm and, knowing that each bundle is long (2-3) microns, we can estimate that for NTPR2 configuration chains of $n_s$=7÷10 and multiple of $n_s$ aligned bundles were connected between the voltage probes. Fig. 2a show a SEM image of two parallel bundles, about 2μm apart, connected to one contact pad while the area of the square with the white perimeter is a zoom of a portion of sample between the probes.

The main panel of Fig.2b shows the *I-V* curves of a sample in the NTPR2 electrode configuration. The change in the slope is similar to that observed for the single junction but it is present at higher voltage values and scales with the electrodes distance. The current value defined in the inset of Fig. 1b as the point where the two straight lines fitting the data cross each other is now $I^*\cong 4\mu A$; this value is the same for all the measured distances but it is higher with respect to the case of the single junction shown in Fig.1b. The lines in the figure represent the best fit of the data in the two current bias regime using the same power law function as in Fig.1b with the same exponents. These *I-V* dependences can be understood by schematically modelling the bundles deposited between two consecutive voltage electrodes as a system of $n_s$ series and $N_p$ parallel resistances (see sketch in the lower inset of Fig.2b). Assuming $n_s$=8, we obtain, by the measured values of the resistance $R_{eq}$=223kΩ between the electrodes spaced at distance $L$=20μm, $N_p$=53 as the number of parallel bundles chains. This model is consistent with the SEM analysis, which show an average distance of 2μm between two nearest bundles connected perpendicular to the 100μm length electrodes. We note in fact that the higher current value $I^* \simeq 4\mu A$ is about 40 times the $I^*$ value measured in the case of the single junction of Fig.1a (0.1 μA): the scaling of the current is of the same order of magnitude of that predictable from the modeling (from which we expect a factor 53). Finally, increasing the distance between the electrodes, a number of $m \cdot n_s$ (with $m$=1,...4) resistances are added in series whereas the number of parallel $N_p$ chains remains the same allowing the same current and different voltages measured in the different electrodes distance in the NTPR2 configuration.

We are now able to estimate the self heating of the sample due to the bias current. This would correspond to the heating of a single bundle in the array of $N_p$=53 parallel bundles. It has been evaluated using the expression[8] $T(x) = T_0 + (p'/g)[1 - \cosh(x/L)/\cosh(L/2L_H)]$ where $T_0$ is the temperature at the contacts, $x$ is the coordinate along the tube, $L$ its length, $p' = I^2 R_{eq} / N_p L$ is the electric power per unit length in the tube, $g$ is the thermal conductance per unit length between the bundles and the substrate and $L_H = (kA/g)^{1/2}$, with $k$ the thermal conductivity and $A$ the cross sectional area of the sample, is the characteristic thermal length of the bundles. The current in each bundle is given by the measured bias current $I$ in the sample divided by the number $N_p$ of parallel bundles and $N_pR_{eq}$ is the resistance of each bundle. Assuming $T_0$=5K, $g$=0.14÷0.2WK$^{-1}$m$^{-1}$ [8], $k$=4000WK$^{-1}$m$^{-1}$[9] as typical values and substituting the measured values for the other parameters, we obtain a thermal heating at the center of the sample of 3μK with respect to the contacts and substrate surface in the case of the highest bias current used ($I$=10μA). This value would correspond to a decrease of the sample resistance less than the experimental uncertainty, and cannot be held responsible for the change in the slope observed between the low and high bias part of the *I-V* curves of Fig.1b.

In Fig.2c we show the temperature dependence of the current-voltage characteristics of the aggregates recorded probing, in the NTPR2 configuration, the voltage between the electrodes spaced 20μm apart. Here we can clearly see the effect mentioned above, namely that the nonlinearity of the current-voltage curves starts below 80K. We can clearly see that decreasing the temperature the “asymptotic” resistance, namely the resistance measured for high voltages decreases as well.

The different slopes in the *I-V* characteristics indicate a non Ohmic behavior of the transport mechanism in the CNT which is more evident at higher bias current where the difference with the expected Ohmic exponent ($\alpha$=1) increases. In order to test this non Ohmic behavior we calculate the resistance from the slope of the *I-V* characteristics for all the voltage electrode distances both in the low and in the high bias current regime. The proof for Ohm's law to be respected is a linear dependence of the electrical resistance on the length of the conductor. Fig. 3a and Fig. 3b show the *R* vs. *L* dependence for all the investigated temperature. The data are plotted in a semi logarithmic scale which is the only way to obtain their rectification.

The results of Fig. 3 demonstrate the non ohmic behavior of our bundles chains. Failures to follow Ohm's law are also observed in isolated SWCNT [10,11] and are attributed to the presence of elastic scattering between electrons and phonons or between electrons and sample impurities[10] or to the presence of anomalous weak localization mechanism[11]. In spite of the somewhat controversial interpretation of the phenomenon, our data clearly show a non Ohmic behavior both in the case of single junction and in the case of the of parallel/series combination of bundles. The lines in Fig.3a and Fig. 3b represent fits to the data acquired at $T$=4K using the phenomenological expression $R = A(\ln(L/L_0)+1)$ where $A$ and $L_0$ are parameters. For both the cases $L_0$ represents the length beyond which the data follow the logarithmic dependence. The fits to the data have been performed for all the measured temperature and the extracted values of $L_0$ are reported as a function of $T$ in Fig. 3c. Both in the case of low and high bias current $L_0$ decreases with increasing the temperature. This behavior is compatible with the increasing of the scattering of electrons inside the sample when the temperature is increased giving rise to a non ohmic behavior on shorter distances.

It is worth noting in the $L_0$ vs. $T$ plot that $L_0$ is almost constant for $T>35K$ while at lower temperature $L_0$ is strictly a decreasing function of $T$. This effect is due to the resistance behavior which, at least at low bias current, sharply increases for temperature below 40K as shown in the inset of Fig.3c where the sheet resistivity as obtained from the data in Fig. 3a and Fig. 3b for $L$=40μm is reported for both low and high bias currents. The data in the inset of Fig.3c are well fitted by FIT model indicating that the junctions formed between the different bundles aligned between the electrodes play the same role as in the case of the single junction shown in Fig.1.

In conclusion, we have reported on the macroscopic investigation of tunneling barrier effects in aggregates of SWCNT bundles. We have provided experimental evidence that the transport properties of the aggregates do not follow Ohmic behavior, however, we have also shown how the properties of the aggregates scale with the length of the aggregates, namely with the number of junctions between the bundles contacting the electrodes. The current-voltage characteristics of the samples can be well understood in terms a simple electrical model accounting for series-parallel connections of the junctions present in the nanotube bundles chains contacting the probe electrodes. Our results show that even a complex systems such as a nanotube bundles aggregate might have solid properties to be considered for potential applications; it is natural to candidate the investigated aggregates as elements with reproducible electrical characteristics for interconnect applications [12], however, the evidence of a temperature-induced nonlinearity in the current voltage singularities it is also a stimulating result in terms of devices physics and sensors.

## Figure Captions

Figure 1. a) SEM image showing a two bundles chain for NTPR1 contact pattern with the arrow indicating the position of the junction; inset: schematic of the four leads configuration. b) the current voltage characteristics measured for the sample shown in a); upper inset: *R*-*T* measurements for the sample shown in a), the line is a fit to the data following the FIT model; lower inset: same data shown in the main panel but in double logarithmic scale. The fit are obtained by power law with different exponents in the different regimes. Also indicated is the method used to determine $I^*$ and $V^*$.

Figure 2. a) SEM image showing two parallel chains departing from the Au contact pad for NTPR2. The inset shows the bundles in a region between the electrodes. The upper inset show an optical image of the current external electrodes and the inner voltage electrodes spaced at distances of 20 microns each. b) The current-voltage characteristics of the aggregate measured using the different voltage electrodes. The fits to the data are obtained by using the same expression of Fig.1b; the dashed line indicates that the change in the slope happens at the same current value for the different electrode distance. Inset: sketch of the series/parallel resistance model assumed for the aggregate deposited between two electrodes. c) *I*-*V* curves at different temperature in the case of $L$=20μm.

Figure 3. a) Differential resistance vs. $L$ dependence measured by the *I*-*V* curves acquired for all the electrode configuration at different temperature and in the regime of low bias current. b) same as a) but in the high bias current regime. c) Temperature dependence of the $L_0$ parameter for low and high bias current. Inset: sheet resistivity per unit length for low and high bias currents. The curves are fit to the data following the FIT model.

.

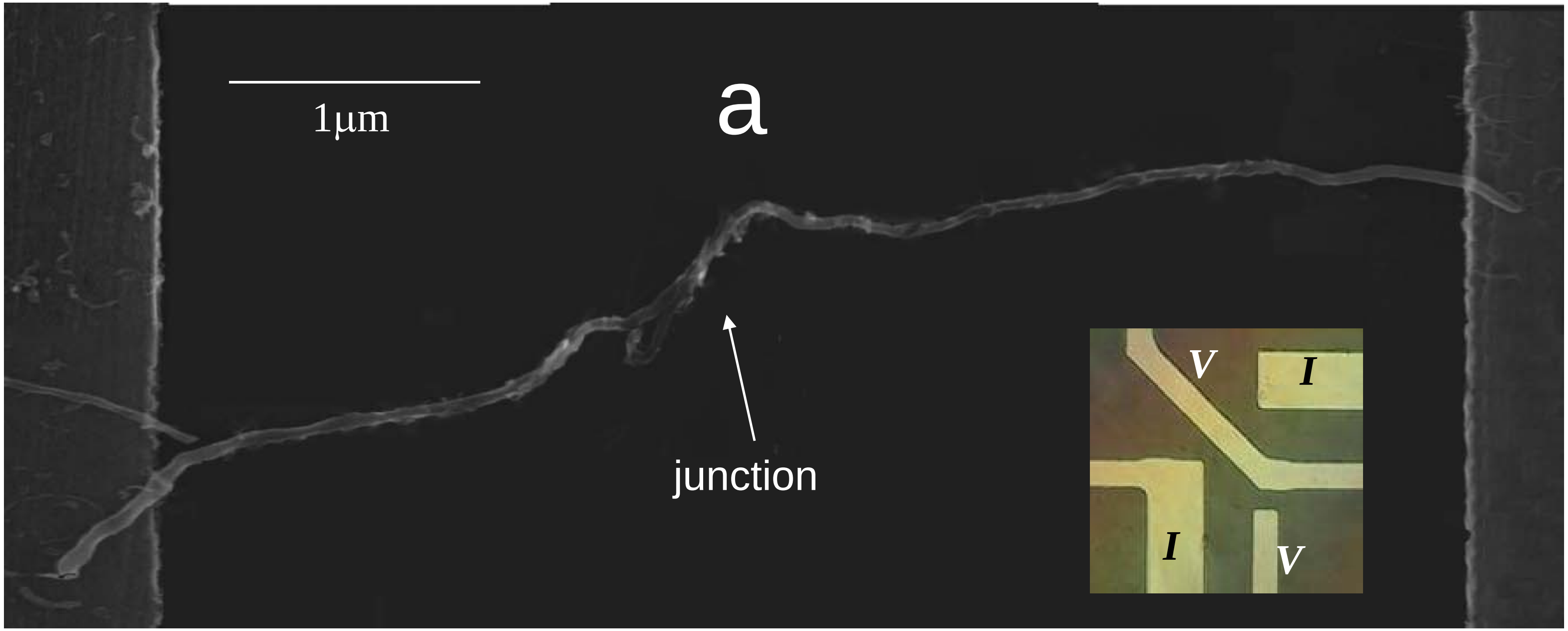

Fig.1a

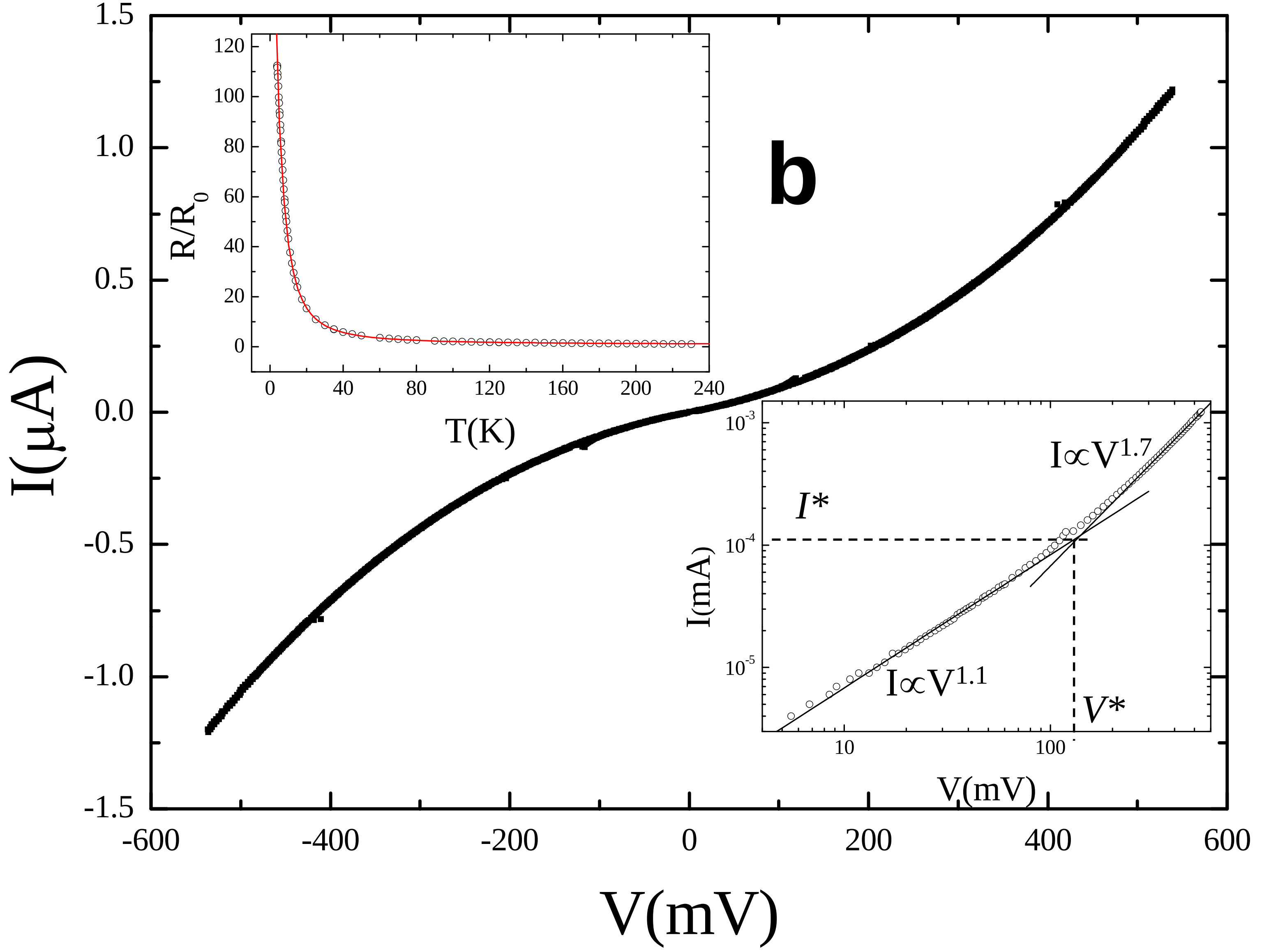

Fig.1b

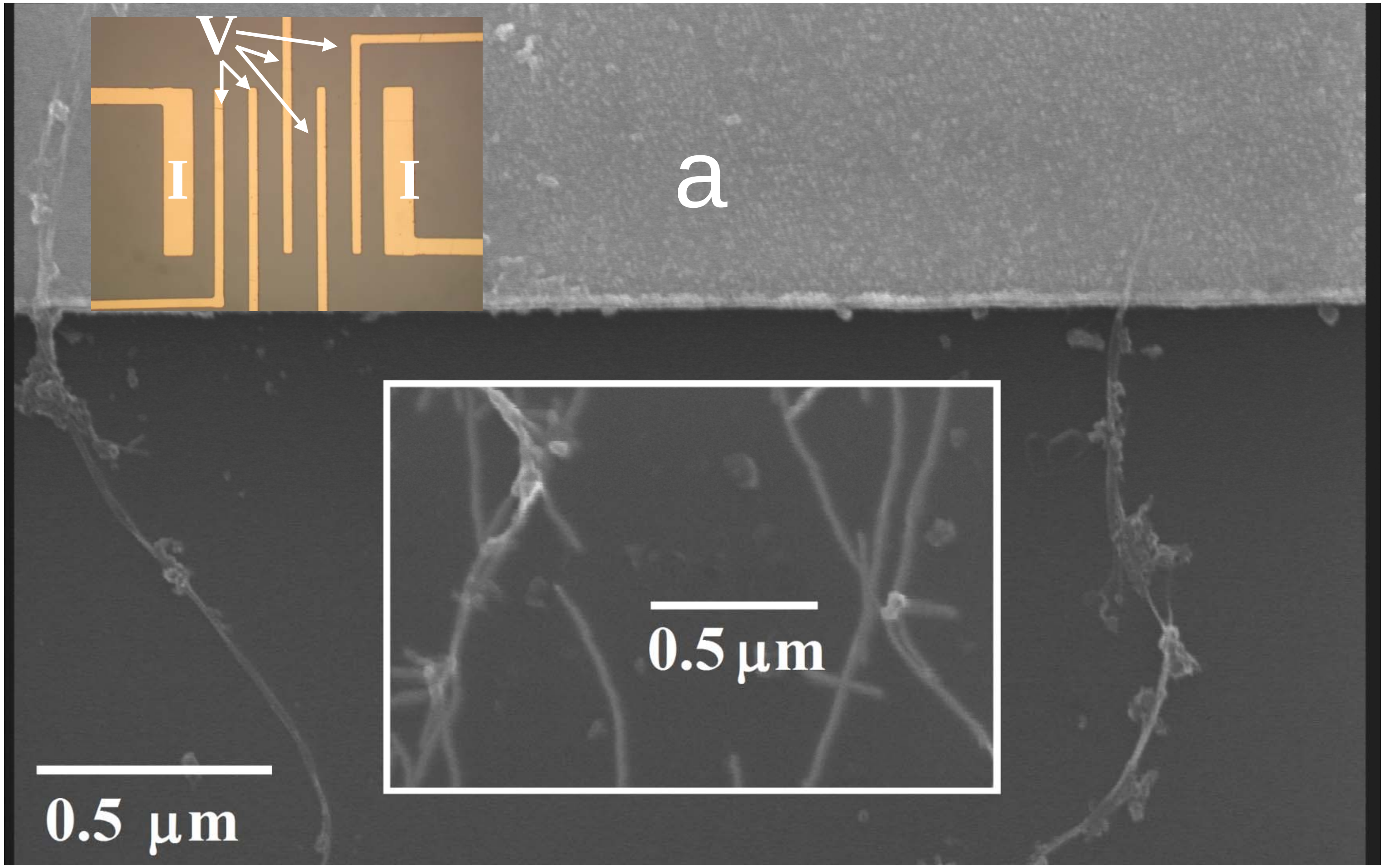

Fig.2a

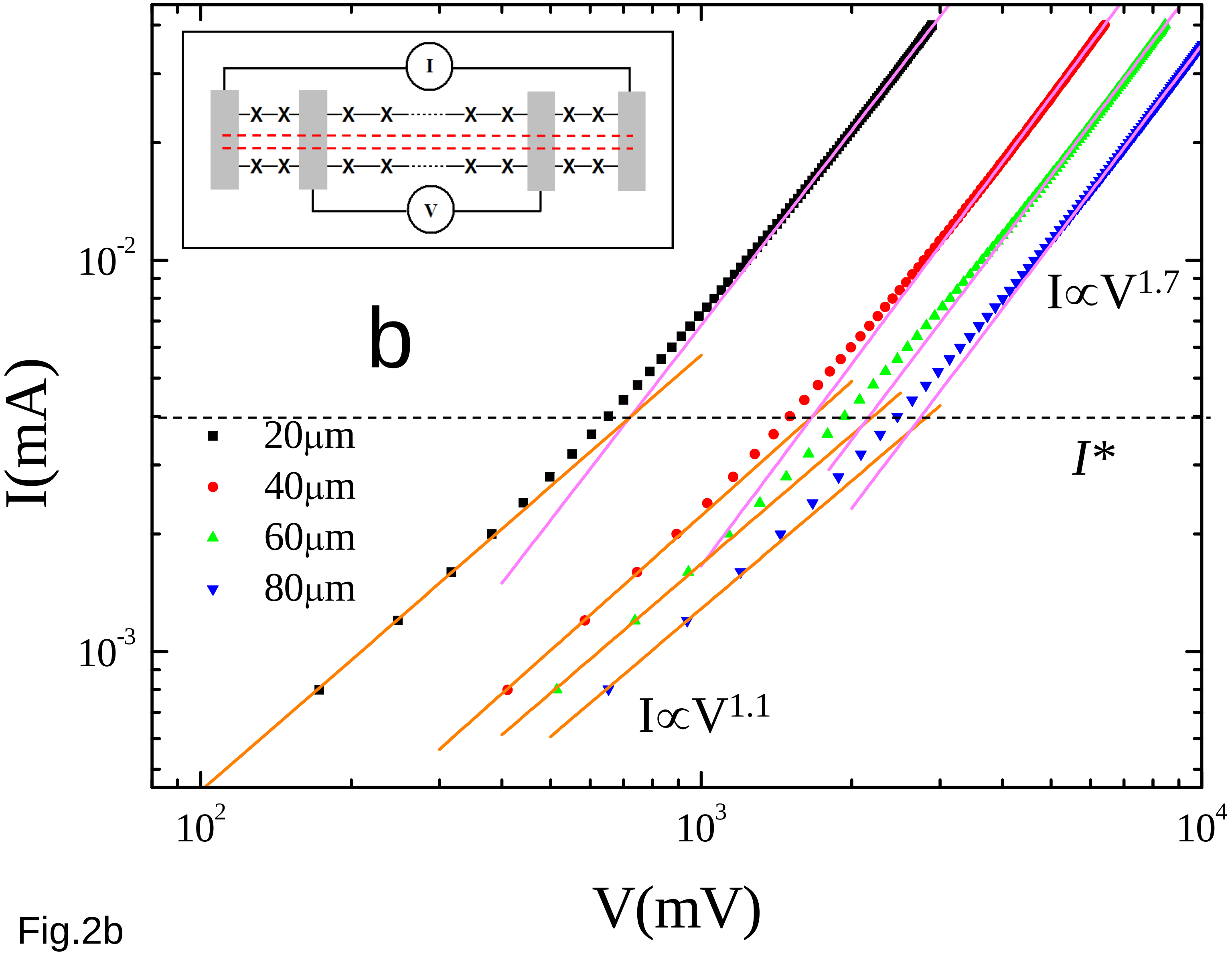

Fig.2b

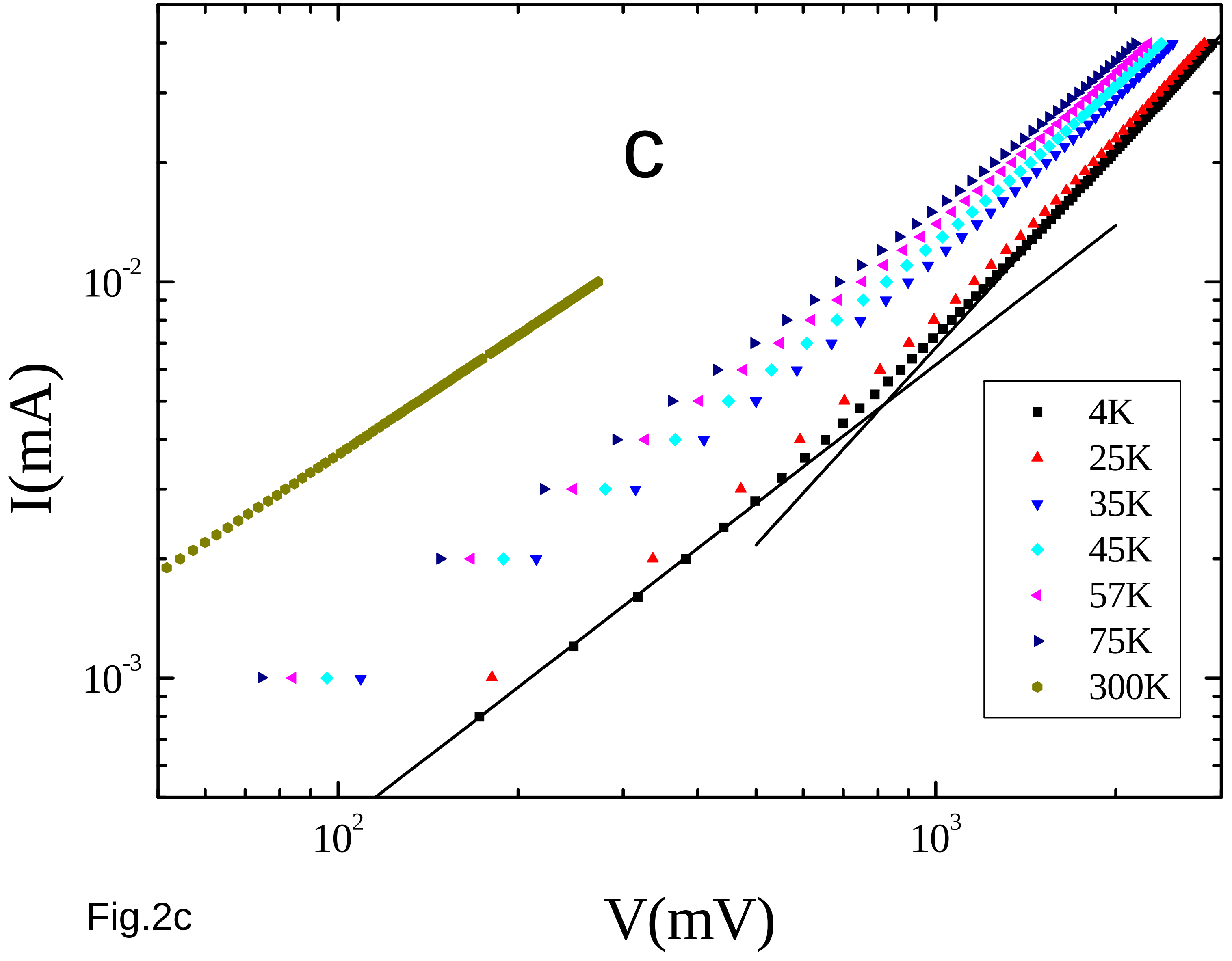

Fig.2c

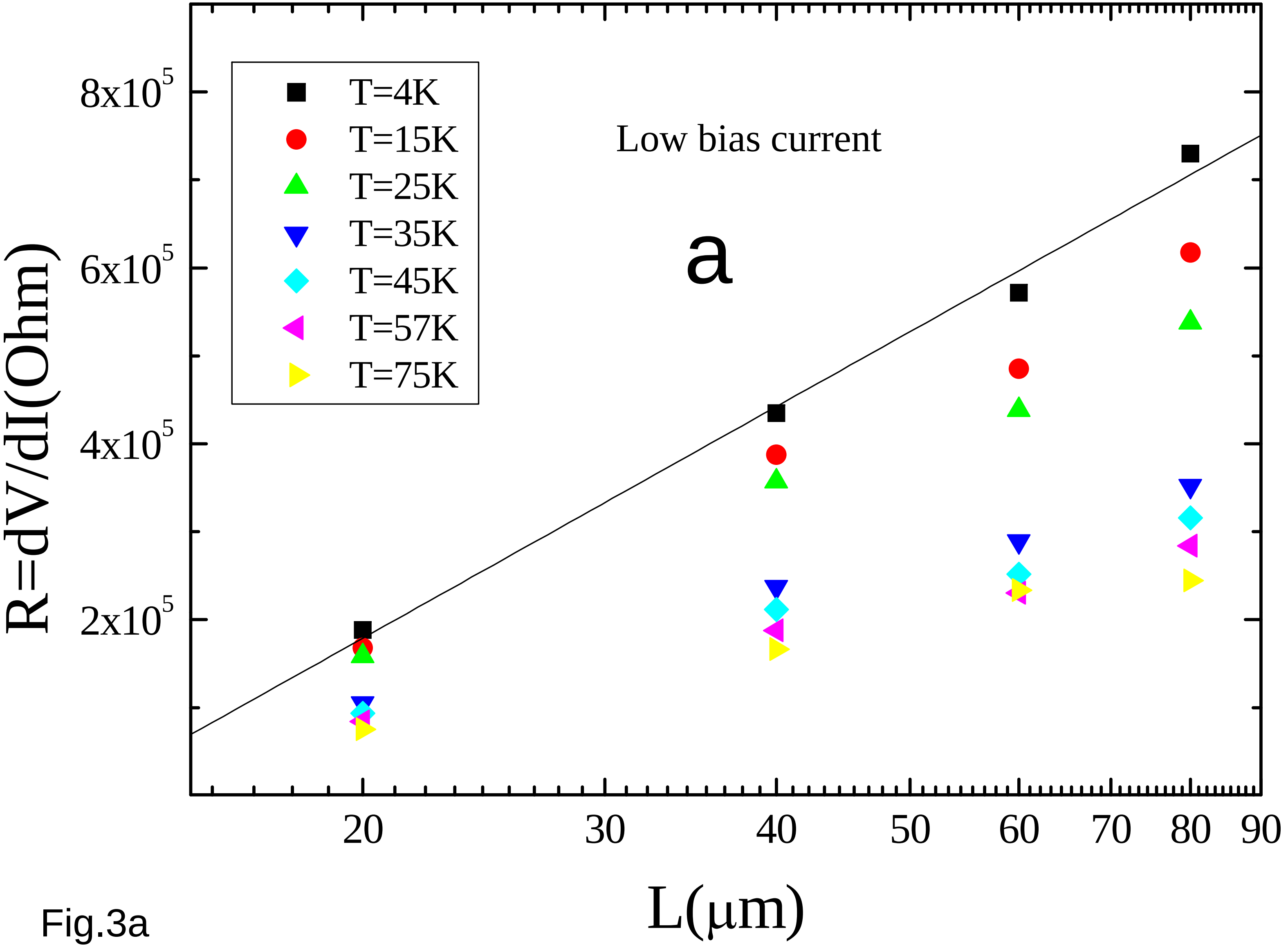

Fig.3a

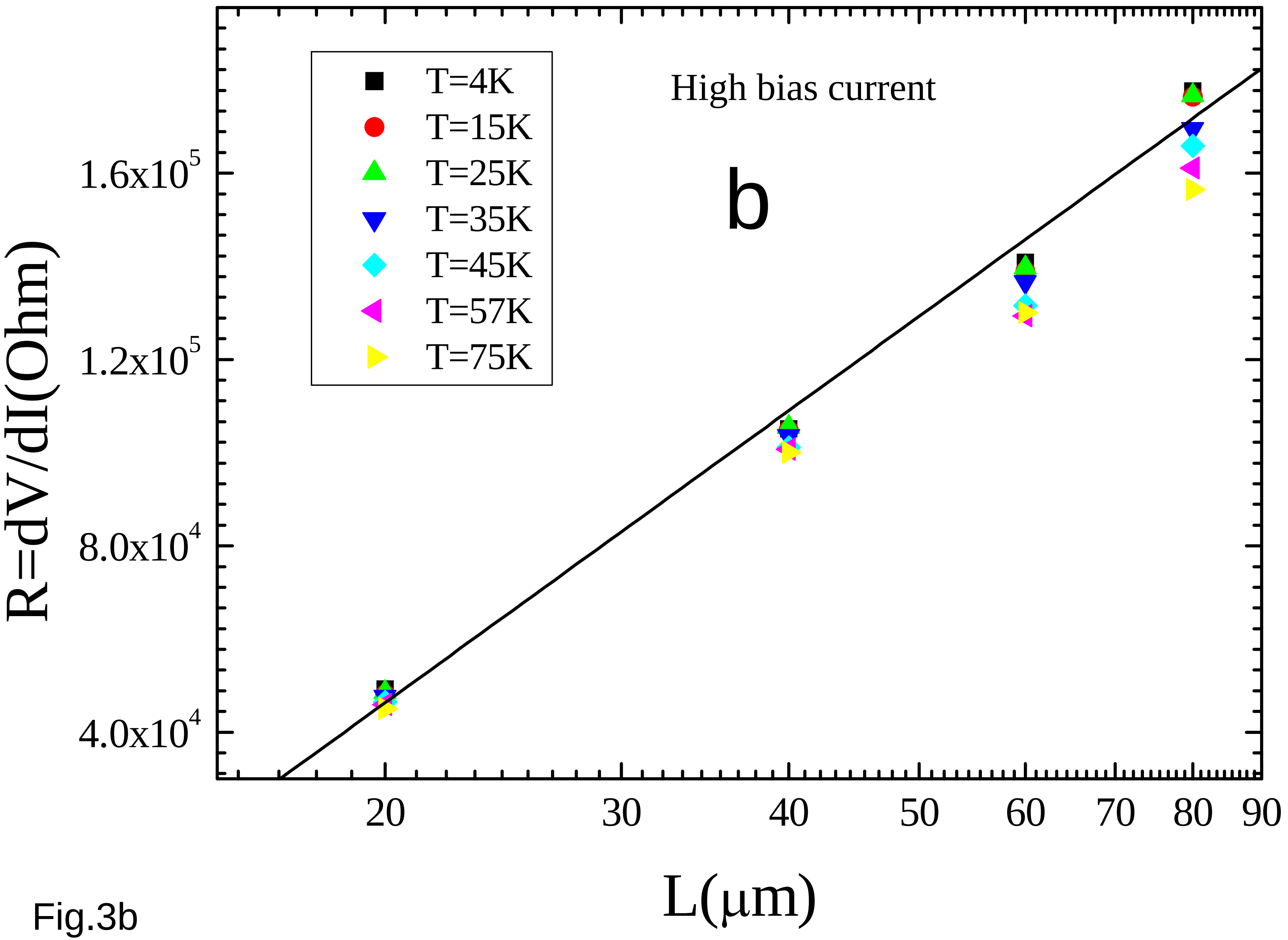

Fig.3b

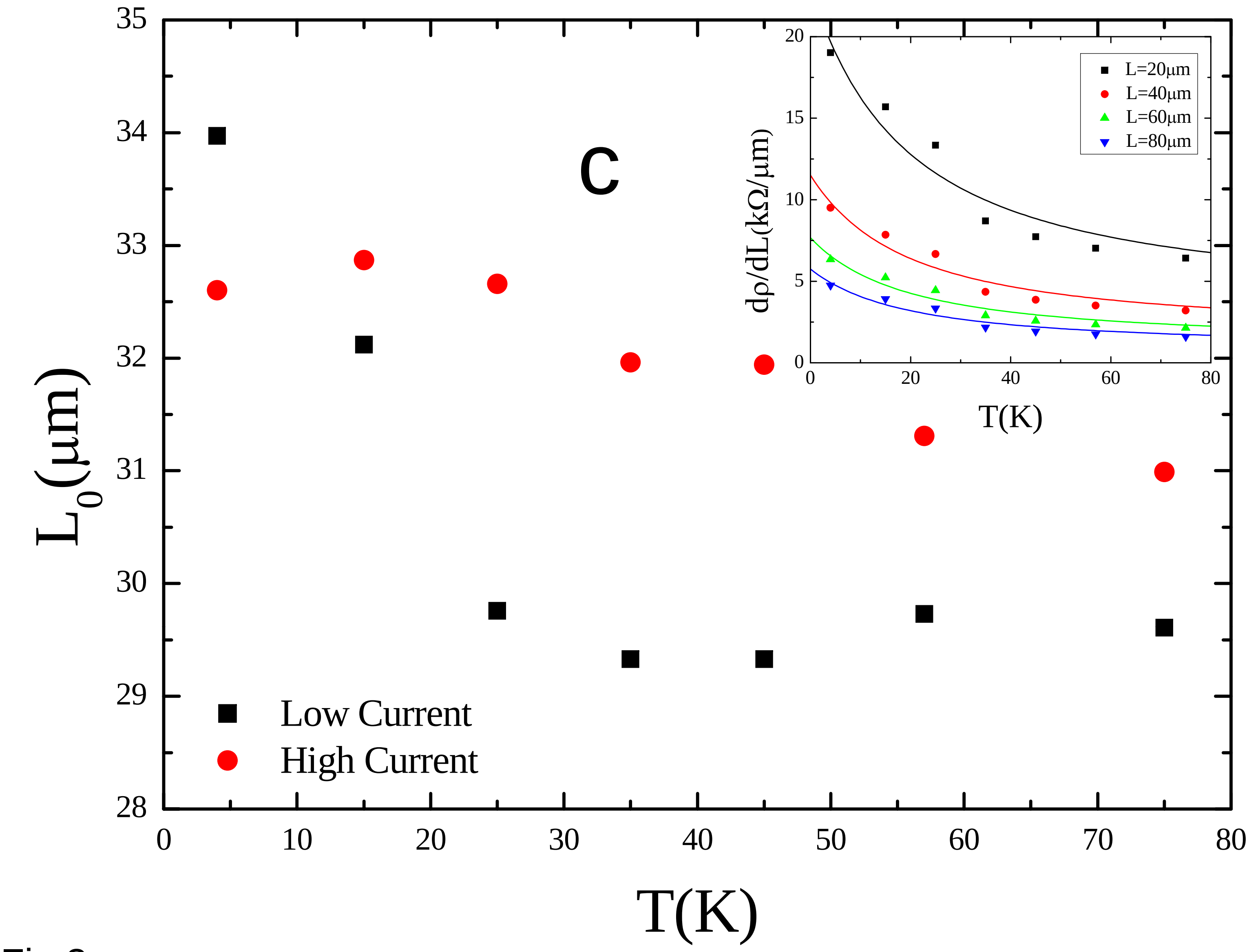

Fig.3c